\documentclass[prl,amsmath,amssymb,aps,twocolumn,superscriptaddress,longbibliography]{revtex4-2}
\usepackage{amssymb}
\usepackage{amsmath}
\usepackage{commath}
\usepackage{verbatim}
\usepackage{graphicx}
\usepackage{dcolumn}
\usepackage{bm}
\usepackage[dvipsnames]{xcolor}
\usepackage[colorlinks]{hyperref}
\usepackage{graphicx}
\usepackage{array}
\usepackage{physics}
\newcolumntype{o}{@{}>{{}}c<{{}}@{}}
\hypersetup{                
	colorlinks=true,                
	breaklinks=true,                
	urlcolor= blue,                
	linkcolor= red,                
    bookmarksopen=false,
	filecolor=black,
	citecolor=blue,
	linkbordercolor=black
}

\begin{document}


\title{Stochastic synthesis-degradation processes: first-passage properties and connections with resetting}

\author{Gabriel Mercado-V\'asquez}
 \email{gabrielmv.fisica@gmail.com}
\affiliation{Pritzker School of Molecular Engineering, University of Chicago, Chicago, IL 60637, USA}%
\author{Denis Boyer}
 \email{boyer@fisica.unam.mx}
\affiliation{Instituto de Física, Universidad Nacional Autónoma de México, Mexico City 04510, Mexico}

\date{\today}

\begin{abstract}
Processes controlled by stochastic synthesis and degradation (SSD) are widespread in biology but their reaction kinetics are not well understood. Using methods borrowed from the theory of resetting processes, we determine the first-passage properties of a collection of independent particles that are synthesized and degraded at constant rates, and follow an arbitrary diffusive motion. At equal synthesis and degradation rates, the mean reaction time with a target site can be minimized as in stochastic resetting, and a $CV$-criterion is derived. When the degradation rate is held fixed and the synthesis costs are taken into account, an optimal synthesis rate is obtained. In bounded domains, despite particle degradation, SSD improves the mean search time compared to a single non-degrading particle if the synthesis rate exceeds a critical value which obeys a universal relation. We illustrate these findings with Brownian diffusion on the infinite line and in an interval.
\end{abstract}

\maketitle

The synthesis and degradation of proteins and other molecular structures is ubiquitous in cellular and intercellular processes. From the replication of simple viruses \cite{Yin1992-ne} to quorum detection in bacteria communication \cite{Waters_2005_quorumsensing}, or pattern formation in complex multi-cellular organisms \cite{Gregor2007-pi,LANDGE20202}, stochastic synthesis-degradation (SSD)  processes are not only widespread but also crucial for robust biological function. 

In the immune system, cell-to-cell communication is mediated primarily by secreted cytokines and chemokines that rapidly diffuse through the extracellular space, enabling propagation of immune signals \cite{Su2024-kj,Altan-Bonnet2019-pl}. Upon signal detection, transcription factors are translocated to the nucleus to initiate RNA transcription, which is subsequently translated into thousands of functional protein copies \cite{Friedman_2006,Yu_2006_ProteinBurst}. However, although hundreds of protein molecules can be synthesized within minutes and production can be sustained for hours \cite{Han2012-xn,Jovanovic_2015,Schwanhausser2011}, protein degradation is tightly regulated in the cell. Due to lysosomal proteolysis and ubiquitination followed by proteasomal degradation, protein half-lives can range from minutes to many hours depending on cellular context \cite{Ross2020ProteomeTI,Thurley_CytokineModelsPlos,Jovanovic_2015,MARTINPEREZ2017283}. Moreover, consumption by surrounding cells can substantially limit the spatial range of cytokine diffusion, restricting it to only a few tens of cell radii \cite{Thurley_CytokineModelsPlos,Altan-Bonnet2019-pl,Oyler-Yaniv2017-ie}. Together, these constraints regulate protein abundance both temporally and spatially within the system \cite{Altan-Bonnet2019-pl}. 

The mechanism of synthesis and degradation of diffusing entities naturally produces concentration gradients and maintains non-equilibrium steady states of the protein profiles in the cell environment \cite{Oyler-Yaniv2017-ie}.
The efficient synthesis and detection of proteins is crucial for maintaining homeostasis and proper physiological function \cite{Su2024-kj,Mosser2021-mi}, for example in the resolution of inflammation \cite{Mosser2008-vr,Murray2011-fh,Akira2006-co} or to inhibit hyperglycemia \cite{Netea2017-ha}.


In living organisms, however,  stochasticity plays a fundamental role in the variability of RNA and protein expression \cite{Ellowitz_2002,Friedman_2006}. In this context, where the main functions of a system are controlled by the arrival of diffusive but degradative molecules at a specific location, understanding first-passage time (FPT) properties is crucial for quantifying reaction efficiency and signal fidelity \cite{Kondev_undated-ih}.

\begin{figure}[t]
    \includegraphics[width=.52\textwidth]{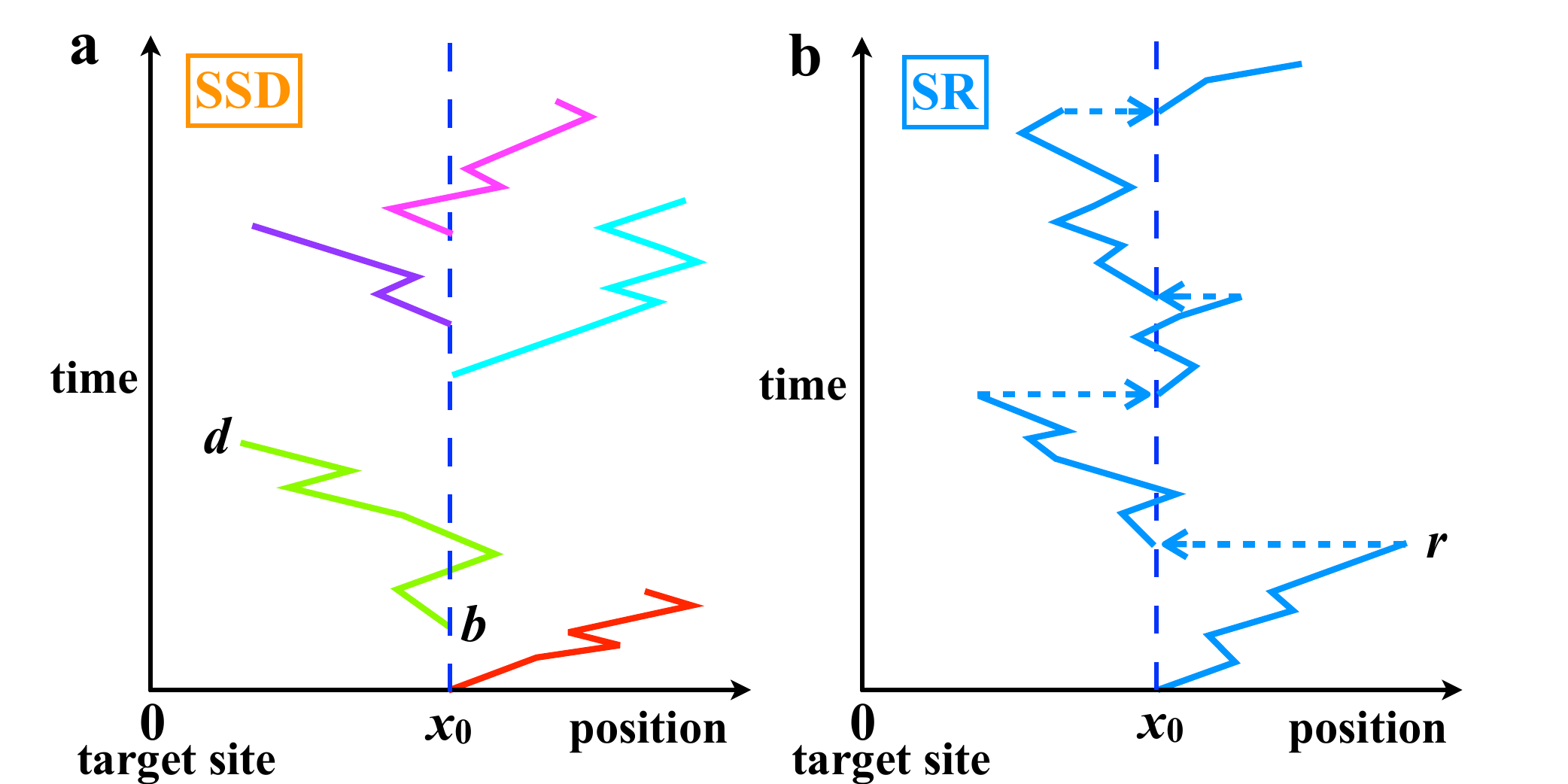}
    \caption{a) In SSD, independent particles (shown with different colors) following an arbitrary stochastic motion are synthesized with rate $b$ at $x_0$ and degrade with rate $d$. b) In SR, a single particle is instantaneously reset to its starting position with rate $r$. Even when $b=d=r$, the two processes have different first-passage statistics, although they exhibit identical densities.}
    \label{fig:numerical}
\end{figure}

A basic SSD model considers independent Brownian particles that are sequentially born at a constant rate $b$ from a fixed position $x_0$. Each particle degrades at a rate $d$ while diffusing, losing its ability to react with \textcolor{blue}{a} target site (Fig. \ref{fig:numerical}a). The ensemble-averaged density of viable particles evolves in the bulk according to a modified diffusion equation \cite{drocco2012synthesis},
\begin{equation}\label{diff}
\frac{\partial \rho(x,t)}{\partial t}=D\Delta\rho(x,t)-d\rho(x,t)+b\delta(x-x_0)\, ,
\end{equation}
where the last two terms describe degradation and synthesis, respectively. A key random variable of interest here is the time \textcolor{blue}{$t_1$} when a target site is reached by a non-degraded particle for the first time.

With $b=d=r$, Eq. (\ref{diff}) turns out to be identical to the forward Fokker-Planck equation describing the diffusion of a single particle under stochastic resetting (SR) \cite{evans_diffusion_2011,evans_diffusion_2011-1}, a different kind of non-equilibrium process illustrated by Fig. \ref{diff}b. In SR, the motion of the particle is interrupted at a rate $r$ and restarts instantaneously from $x_0$. Resetting processes are efficient for completing search tasks and they have received a considerable attention in the past decade, with extensions well beyond simple diffusion. Their first-passage statistics are well-understood \cite{evans_stochastic_2020,gupta2022stochastic, kusmierz2014first,kusmierz_optimal_2015,montero2017continuous,evans2018run,chechkin2018random}. Despite its proximity with SR, SSD has in principle its own reaction kinetics, which remain elusive. SR can actually be viewed as a \lq\lq mean-field" SSD process, where death and birth events are perfectly correlated, so that 
the number ${\cal N}(t)$ of particles in the system at time $t$ is always 1 (or some {fixed} other number \cite{biroli2023extreme}). In SSD, ${\cal N}(t)$ fluctuates around 
its average value, which is also time-dependent, $\langle{\cal N}(t)\rangle=\int \dd x\,\rho(x,t)=
b/d+[{\cal N}(t=0)-b/d]e^{-dt}$, and tends to $b/d$.

In this Letter, we adapt methods from resetting theory to derive the first-passage properties of an arbitrary process under SSD, extending previous work with $d=0$ \cite{campos2024dynamics,tung2025first}. First-passage observables can be written in terms of the probability distribution function (PDF) of the FPT of a single non-degrading particle. We determine general conditions under which the mean first-passage time (MFPT) and related quantities can be optimized in open and bounded domains. If synthesis exceeds a critical rate obeying a simple universal relation, SSD processes speed up the reaction compared to that of a single nondegrading particle. We illustrate the findings with a few Brownian examples


{\it Survival probability.} In the setup depicted by Fig. \ref{fig:numerical}a, {two quantities of main interest are the probabilities $Q^{(1)}(t)$ and $Q^{(0)}(t)$, respectively, that no particle has reached the target site during the time interval $[0,t]$, given that the system at $t=0$ was composed of $1$ ($0$, resp.) particle, at position $x_0$. Hence $Q^{(0)}(t)$ is the target survival probability in the absence of searchers initially. Likewise, without synthesis ($b=0$), 
$q_{d}(t)$ denotes the probability that a particle starting from $x_0$ has not reached the target in $[0,t]$. Without synthesis and degradation ($b=d=0$), $P_0(t)$ is the PDF of the FPT at the target site of a particle with initial position $x_0$.}
The dimensionality, geometry, or type of diffusion are arbitrary. 

{
The first particle has not degraded yet at time $\tau$ with probability $e^{-d\tau}$ and
the trajectories that find the target in a time $\tau<t$ do not contribute to $q_d(t)$.
Therefore \cite{meerson2015mortality},}
\begin{eqnarray}\label{qm}
    q_{d}(t)=1-\int_0^t d\tau e^{-d\tau}P_0(\tau)\, . 
\end{eqnarray}
The independence between particles also implies that
\begin{equation}\label{q1q0}
Q^{(1)}(t)=q_d(t)Q^{(0)}(t)\, .
\end{equation}
In Eq. (\ref{q1q0}), the whole SSD process is viewed as composed of the trajectory of the first particle starting at $t=0$ [associated to $q_d(t)$], superposed to the SSD process without an initial particle, representing all the particles synthesized after $t=0$.
The quantity $Q^{(0)}(t)$ obeys an integral, renewal-like equation,
\begin{equation}\label{renewq0}
Q^{(0)}(t)=e^{-bt}+b\int_0^{t}{\rm d} t'\, e^{-b(t-t')}q_d(t')Q^{(0)}(t')\, .
\end{equation}
The first term in the rhs of Eq. (\ref{renewq0}) is the probability that no particle is born during the time interval $[0,t]$. 
The second term describes the probability that the first particle was born before $t$, in the time interval $[t-t',t-t'+{\rm d}t']$, an event that occurs with probability $b{\rm d}t' e^{-b(t-t')}$. 
{In the remaining time interval $[t-t',t]$, the probability  
of not finding the target} is precisely $Q^{(1)}(t')$. Using Eq. (\ref{q1q0}), the second term of Eq. (\ref{renewq0}) is obtained.

Unlike in the renewal approach of stochastic resetting \cite{pal_diffusion_2016, evans2018run,chechkin2018random}, Eq. (\ref{renewq0}) does not have a convolution structure {because when the 2nd particle is born the first one may still be diffusing, implying a new initial condition with two particles. However, Eq. (\ref{renewq0}) is exactly solvable (see \cite{SM}),}
\begin{equation}\label{q0exact0}
Q^{(0)}(t)=e^{-b\int_0^t {\rm d}u[1-q_d(u)]}\, .
\end{equation}
Using Eq. (\ref{qm}) and integrating by parts, this expression can also be written as
\begin{equation}\label{q0exact0b}
Q^{(0)}(t)=e^{-b\int_0^t {\rm d}u\, e^{-du}(t-u)P_0(u)}\, .
\end{equation}
The general expression for $Q^{(1)}$ is thus
\begin{equation}\label{q1exact0}
Q^{(1)}(t)=q_d(t)\, 
{e^{-b\int_0^t {\rm d}u\, e^{-du}(t-u)P_0(u)}\, .
}
\end{equation}
Eqs. (\ref{q0exact0b}) and (\ref{q1exact0}) provide explicit forms of the survival probabilities at all times (if the FPT distribution $P_0(t)$ of the underlying process is known), in contrast with SR 
where the {complete} time dependence is rarely available.

The late time behaviors of $Q^{(0,1)}(t)$ are deduced by noticing that $q_d(t)$ rapidly tends to 
{$1-\widetilde{P}_0(d)<1$ if $t\gg 1/d$, where  $\widetilde{P}_0(d)=\int_0^{\infty}{\rm d}t\, e^{-dt}P_0(t)$ denotes the Laplace transform of $P_0$. The integral upper bound ($t$) can also be replaced by $\infty$ in Eqs. (\ref{q0exact0b})-(\ref{q1exact0}), and}
\begin{equation}\label{Qasympt}
Q^{(i)}(t)\simeq c_i e^{-b\widetilde{P}_0(d)t}\, ,\quad i=0,1
\end{equation}
with the prefactors $c_0=e^{b\int_0^{\infty}{\rm d}u\, e^{-du}u P_0(u)}$ and $c_1=c_0[1-\widetilde{P}_0(d)]$.
{The exponential form
(\ref{Qasympt}) yields a first important quantity related to the search efficiency of SSD processes, i.e., the exact characteristic reaction time}
\begin{equation}\label{taubd}
\tau_{b,d}=1/[b\widetilde{P}_0(d)]\,,
\end{equation}
which is inversely proportional to $b$, while the effect of $d$ depends on the type of diffusion and geometry. Since $\tau_{b,d}<\infty$ for any $b>0$, SSD  shares with SR an advantageous feature: underlying processes having a fat-tailed survival probability, or $q_0(t)\sim t^{-\theta}$ with $0\le\theta<1$ (implying an infinite MFPT), {become short-tailed} and completed within a finite time.

We apply this finding to the 1D Brownian particle on the semi-infinite line with an absorbing target at $x=0$ (see Fig. \ref{fig:numerical}a), where $\theta=1/2$ and $P_0(t)=\frac{|x_0|}{\sqrt{4\pi D}t^{3/2}}e^{-\frac{x_0^2}{4Dt}}$ is the L\'evy-Smirnov distribution, with $\widetilde{P}_0(d)=e^{-\sqrt{d/D}|x_0|}$. From Eq. (\ref{taubd}), $\tau_{b,d}= b^{-1}e^{\sqrt{d/D}|x_0|}$, which diverges when $b\to 0$ (slow synthesis) or $d\to \infty$ (fast degradation).

To compare with the resetting phenomenology, let us set $b=d=r$ and vary $r$. In the above example, $\tau_{r,r}=r^{-1}e^{\sqrt{r/D}|x_0|}$ diverges at both small and large $r$, and has a global minimum at $r^*=4D/x_0^2$. Hence there is an optimal rate for which the no-reaction probability decays the fastest to 0.  
{In Brownian diffusion under SR, the decay of the survival probability is also exponential at large $t$, and the associated characteristic time  $\tau^{(SR)}_r$ is governed by a pole, solution of a  transcendental equation.
At both small and large $r$, $\tau^{(SR)}_r$  has the same diverging behaviors than  $\tau_{r,r}$ above
\cite{evans_diffusion_2011, evans_stochastic_2020}.}

{\it Mean first-passage time and search costs.}
The MFPT denoted as $T_{b,d}$ is obtained from the general relation $T_{b,d}=\int_0^{\infty}{\rm d}t\, Q^{(1)}(t)$. One obtains \cite{SM},
\begin{equation}\label{T1gen}
T_{b,d}=\int_0^{\infty} {\rm d}t\,   e^{-b\int_0^t {\rm d}u\, e^{-du}(t-u)P_0(u)}\,-\frac{1}{b}\,.
\end{equation}
Contrary to its SR counterpart, given by $T_r^{(SR)}=[1-\widetilde{P}_0(r)]/[r\widetilde{P}_0(r)]$ \cite{evans2018run, evans_stochastic_2020}, $T_{b,d}$ cannot be expressed in terms of elementary functions in principle.
{Yet, its numerical evaluation is as fast as for a single integral.}

\begin{figure}[t]
    \includegraphics[width=.5\textwidth]{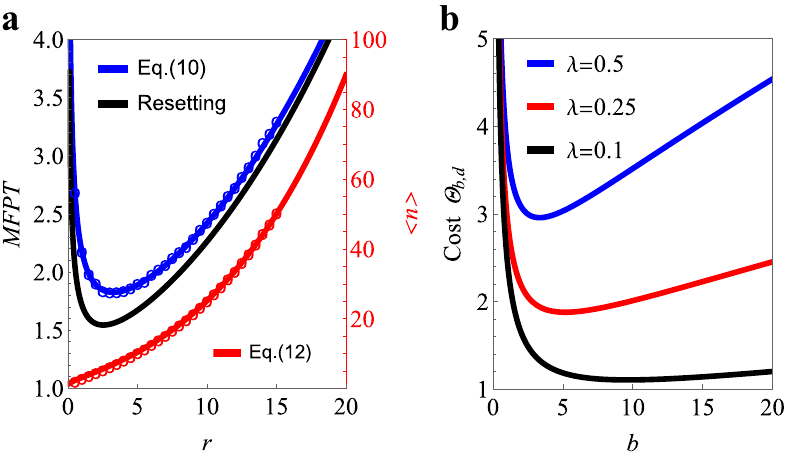}
    \caption{SSD Brownian particles with $D=1$ starting from $x_0=1$ on the semi-infinite line with an absorbing boundary at $x=0$. a) MFPT in Eq. (\ref{T1gen}) (blue line) and average number of synthesized particles until absorption (red line, right y-axis) with $b=d=r$. For comparison we depict the MFPT for stochastic resetting ({black line}) \cite{evans_diffusion_2011}. The symbols represent Langevin dynamics simulations. b) Keeping degradation constant ($d=1$), total cost function as a function of the production rate $b$ for different values of the synthesis cost parameter $\lambda$.}
    \label{fig:cost}
\end{figure}

Fig. \ref{fig:cost}a shows $T_{r,r}$, in the case $b=d=r$ for Brownian particles on the 1D line. Similarly to $\tau_{r,r}$, the MFPT is non-monotonous with $r$ and exhibits a minimum at $r_{\rm min}\simeq 3.256\, D/x_0^2$. $T_{r,r}$ is larger than $T^{(SR)}_{r}$, also displayed, indicating that fluctuations in the number of particles around 1 prolong the reaction time. 

Let us now consider the general case, $b\neq d$. Assuming that $d$ is held constant, the minimization of $\tau_{b,d}$ or $T_{b,d}$ with respect to $b$ yields the optimal value $b=\infty$. However, this cannot be achieved in practice because of the costs associated with large synthesis rates. The costs incurred by SSD processes differ from their SR counterparts. In SSD, degradation is {largely spontaneous and cost-free, while} in resetting, the particle must be brought back from its current location to the starting point by a certain protocol \cite{pal_diffusion_2016, pal2019invariants,tal2020experimental,bodrova2020resetting, olsen2024thermodynamic,gupta2025optimizing,biswas2025target}. Synthesis however requires energy and resources
{(e.g., DNA transcription by molecular motors leading to protein synthesis is fueled by ATP)}, ideally the same amount for each {new particle produced}.

{Therefore, a natural way of evaluating the energetic cost of a search is to assume that it is proportional to} the total number ($n$) of {searchers employed in the time interval $[0,t_1]$, where $t_1$ is the FPT of the reaction. 
The joint distribution of the random variables $n$ and $t_1$ is given by} \cite{SM}
\begin{equation}\label{joint}
P(t_1,n)=\frac{\partial Y_n(t_1)}{\partial t_1}\frac{b^{n-1}}{(n-1)!}e^{-bt_1}\,,
\end{equation}
with $Y_n(t_1)=
[1-q_d(t_1)][\int_0^{t_1} {\rm d}u\,q_d(u)]^{n-1}$, leading to \cite{SM}{
\begin{equation}\label{nav}
\langle n\rangle \equiv\sum_{n=1}^{\infty}n\int_0^{\infty}{\rm d}t_1\,P(t_1,n)=1+bT_{b,d}\,.
\end{equation}}
{The last equality} in Eq. (\ref{nav}) has a simple interpretation. The particles created until the completion time $t_1$ are the initial one plus all the subsequent particles born during $(0,t_1]$, in number $b\langle t_1\rangle=bT_{b,d}$ on average. 
As shown by Fig. \ref{fig:cost}a in the Brownian example with $b=d=r$, $\langle n \rangle$ increases monotonously with $r$ and is about $6.94$ at $r=r_{\rm min}$.

The synthesis costs can be added to the {MFPT (see \cite{biswas2025target} for a similar example) to build} a total cost function
\begin{equation}
\Theta_{b,d}=T_{b,d}+\lambda \langle n\rangle\,,
\end{equation}
{where $\langle n\rangle$ is given by Eq. (\ref{nav})} and
$\lambda$ a parameter with dimension of time. As shown in Fig \ref{fig:cost}b, fixing $d$, $\Theta_{b,d}$ is minimum at a certain $b_{\rm min}$ for any value of $\lambda>0$. As $\lambda$ becomes larger, the overall optimization is more sensitive to the choice of $b$.


{\it Small $b$ and $d$ expansion, CV-criterion.} 
We now assume that $P_0(t)$ has a finite first moment $\langle T_0\rangle$, as observed in bounded domains, and wish to compare the MFPT of a SSD process to this bare search time $\langle T_0\rangle$ (here, $\lambda=0$). To do so, it is instructive to analyze the behavior of $T_{b,d}$ when both $b$ and $d$ are small. This strategy is well-known in resetting, where the study of $T_{r}^{(SR)}$ at small $r$ allows the derivation of a universal relation (the so-called CV-criterion) telling whether resetting expedites search or not for a given process \cite{reuveni_optimal_2016,pal2017first,pal2022inspection,pal2024random}.

{As discussed in \cite{mejia2011first}, the variance of the FPT is relevant to many search problems.}
Let us assume that $P_0(t)$ has a finite variance $\langle T_0^2\rangle-\langle T_0\rangle^2$ and decays exponentially or faster at large $t$, over a time-scale $1/\mu$.
The expansion of $T_{b,d}$ with $b,d\ll\mu$ is somewhat subtle and leads to the general expression \cite{SM}
\begin{equation}\label{T1pert}
T_{b,d}=\frac{b+d}{b}\langle T_0\rangle-\frac{(b+d)^2}{2b} \langle T_0\rangle^2 \left[CV_0^2-\frac{d^2}{(b+d)^2}\right]+ {\rm h.o.t.}\,,
\end{equation}
where $CV_0=\sqrt{\langle T_0^2\rangle-\langle T_0\rangle^2}/\langle T_0\rangle$ is the coefficient of variation of the FPT of the underlying process, while h.o.t. represents negligible terms of order $b^2$, $d^2$ or $db$.

When $b=d=r$, the above relation becomes
\begin{equation}\label{CV}
T_{r,r}=2\langle T_0\rangle-2 r \langle T_0\rangle^2 \left [CV_0^2-\frac{1}{4}\right] + {\cal O}(r^2)\, ,
\end{equation}
instead of 
\begin{equation}\label{CVreset}
T_{r}^{(SR)}= \langle T_0\rangle-\frac{r}{2} \langle T_0\rangle^2 [CV_0^2-1] + {\cal O}(r^2)\, ,
\end{equation}
for resetting processes \cite{pal2024random}.
One can notice two fundamental differences between SSD and SR. 

\begin{figure}[t]
    \includegraphics[width=.5\textwidth]{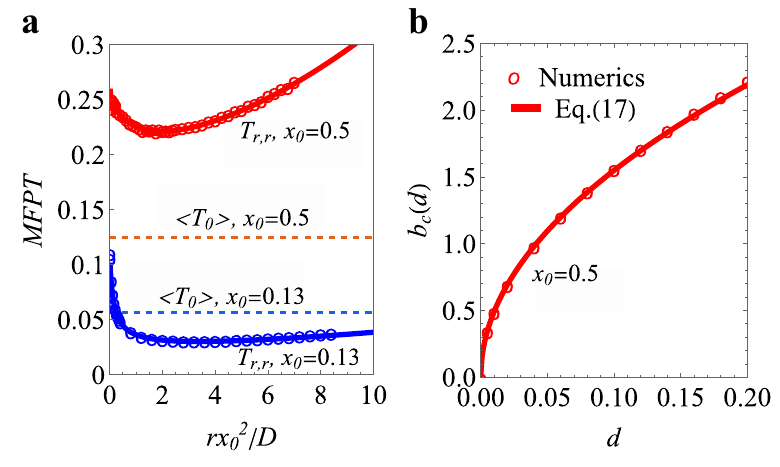}
    \caption{Brownian particles with $D=1$ in the interval $[0,1]$. a) MFPT outside of the interval with  $b=d=r$. The solid lines are obtained from Eq. (\ref{T1gen}), see \cite{SM} for details, and the symbols from Langevin dynamics simulations. The horizontal dotted lines represent the mean exit time $\langle T_0\rangle$ for a single non-degrading particle ($r=0$). b) Scaling-law (\ref{bcritdsmall}) for the critical synthesis rate vs. $d$ at fixed $x_0$. The symbols are numerical solutions of $T_{b,d}=\langle T_0\rangle$ with $T_{b,d}$ given by Eq. (\ref{T1gen}).}
    \label{fig:MFPT}
\end{figure}

Firstly, $\lim_{r\to0}T_{r,r}= 2\langle T_0\rangle$ and not $\langle T_0\rangle$, which may look surprising at first sight. The factor 2 (more generally $\frac{b+d}{b}$) can be explained qualitatively as follows. At small $r$, the first particle diffuses alone and reacts in an average time $\langle T_0\rangle$, assuming that it has not degraded before. The probability of not degrading is roughly $e^{-r\langle T_0\rangle}$, close to 1. With the small complementary probability $1-e^{-r\langle T_0\rangle}$, the first particle dies before locating the target and one needs to wait for a very long time ($1/r$ on average) before the second particle is created. From that instant, this particle will take a time $\langle T_0\rangle$ to locate the target. Taking the average over these two search times gives
$
\langle T_0\rangle e^{-r\langle T_0\rangle}+\left(r^{-1}+ \langle T_0\rangle\right)(1-e^{-r\langle T_0\rangle})\to 2\langle T_0\rangle\, ,
$
as $r\to0$. This limit (implying a discontinuity at $r=0$) is confirmed by Langevin dynamics simulations of 1D Brownian particles in the interval $[0,1]$ with absorbing boundaries in $x=0$ and $1$, see Fig. \ref{fig:MFPT}a. 

Secondly, $T_{r,r}$ decreases with $r$ at small $r$ from the value $2\langle T_0\rangle$ if 
$CV_0>\frac{1}{2}$, see Eq. (\ref{CV}).
In that case, there exists in principle an non-trivial rate $r_{\rm min}$ where $T_{r,r}$ has a local minimum, like in the semi-infinite domain. Conversely, if $CV_0<1/2$, the mean time $T_{r,r}$ increases with $r$. The threshold value of $1/2$, instead of $1$ for stochastic resetting in Eq. (\ref{CVreset}), implies that a much wider class of processes will exhibit a non-monotonous behavior when subject to SSD. Processes that exhibit a monotonous increase of the MFPT with $r$ under SR may have a non-monotonous behavior under SSD. 

In the above example of Brownian motion in the interval, the smallest $CV_0$ is achieved by choosing the starting position at the middle of the domain ($x_0=1/2$), and the corresponding value is $\sqrt{2/3}$ (see e.g., \cite{dagdug_diffusion_2024}). As $\sqrt{2/3}<1$, resetting increases the exit time \cite{pal2019first}. But since $\sqrt{2/3}>1/2$, the MFPT decreases with $r$ at small $r$ under SSD, see Fig. \ref{fig:MFPT}a. Unlike in resetting \cite{pal2019first}, a non-trivial $r_{\rm min}$ locally minimizes $T_{r,r}$ for any initial position $x_0$ in the interval. Interestingly, as also shown by Fig. \ref{fig:MFPT}a, the minimum of $T_{r,r}$ at $r_{\rm min}$ can be larger than $\langle T_0\rangle$ (for $x_0=0.5$), or smaller ($x_0=0.13$). In the latter case, SSD with $b=d$ turns out to be beneficial for search.


{\it Critical synthesis rate.} The above findings lead us to a general question in bounded domains. Given a process with a prescribed degradation rate $d$, what is the synthesis rate $b_c(d)$ above which the MFPT $T_{b,d}$ becomes {\em lower} than $\langle T_0\rangle$ in the same setting? In other words, how the degradation of the individual searchers can be compensated through the synthesis of more searchers, in order to achieve identical or faster search times than a single particle with $d=0$?

In the case $d=0$, Eq. (\ref{T1pert}) gives
$T_{b,0}=\langle T_0\rangle-\frac{b}{2}\langle T_0\rangle^2CV_0^2+{\cal O}(b^2)$. Clearly, any non-zero synthesis rate improves the MFPT of the first particle alone, or $b_c(d=0)=0$. 
If $d$ is non-zero, a minimal synthesis rate is necessary to improve $\langle T_0\rangle$. At small $d$, we assume that $b_c(d)$ is also small, so that Eq. (\ref{T1pert}) can be used. Solving $T_{b,d}=\langle T_0\rangle$ for $b$, a universal relation is obtained
\begin{equation}\label{bcritdsmall}
b_c(d)=\frac{1}{CV_0}\sqrt{\frac{2d}{\langle T_0\rangle}}+{\cal O}(d)\,.
\end{equation}
From Eq. (\ref{bcritdsmall}), the critical synthesis rate is much larger than $d$ in the small $d$ limit. Both the square-root dependence and the prefactor are completely general, as Eq. (\ref{T1pert}) is independent of the geometry, spatial dimension or type of diffusion. 
The scaling relation (\ref{bcritdsmall}) agrees very well with the $b_c(d)$ obtained numerically in the example of Brownian particles in an interval (see Fig. \ref{fig:MFPT}b).

{\it Conclusion.} Inspired from methods used in the study of resetting processes, we have analyzed the first-passage properties of  stochastic synthesis-degradation (SSD) processes. The results are expressed in terms of the first-passage time PDF of the underlying process in the absence of SSD, and they are not limited to Brownian diffusion or even Markov processes. SSD can reduce the completion times of target search tasks in bounded and unbounded domains, and they can be optimized. 

{Protein synthesis often occurs in bursts \cite{Pedraza_2008,Kumar_2015,Friedman_2006}, with nonexponential waiting times between transcription events, as observed in both bacterial and mammalian gene regulation \cite{Harper_2011,Taniguchi_2010}. 
This bursty production increases variability in protein levels \cite{Pedraza_2008,Friedman_2006}, which is especially critical at low concentrations, where diffusion-limited binding can strongly influence first-passage time properties \cite{Pulkkinen_2013}. 
The renewal-based framework employed here is sufficiently flexible to accommodate such non-Poissonian synthesis–degradation kinetics}.

During the preparation of this manuscript, we became aware of a recent preprint presenting some of the results reported here independently \cite{linn2026dynamic}.
DB acknowledges support from Ciencia de Frontera 2019 Grant 10872 (Conacyt, M\'exico).







%

\onecolumngrid

\newpage

\begin{center}
{\Large Supplemental Material of\\
 $\ $\\
\lq\lq Stochastic synthesis-degradation processes: first passage properties and connections with resetting"}\\
$\ $\\
G. Mercado-V\'asquez and D. Boyer
\end{center}

\vspace{1cm}

\setcounter{equation}{0}

\section{Calculation of the survival probability $Q^{(0)}(t)$}\label{appA}

The survival probability $Q^{(0)}(t)$ obeys the renewal-like relation
\begin{equation}\label{renewq0A}
Q^{(0)}(t)=e^{-bt}+b\int_0^{t}{\rm d} t'\, e^{-b(t-t')}q_d(t')Q^{(0)}(t')\, ,
\end{equation}
with 
\begin{equation}\label{qdA}
q_d(t)=1-\int_0^t{\rm d}u\,P_0(u)e^{-ru}.
\end{equation}
Eq. (\ref{renewq0A}) can be rewritten as
\begin{equation}
Q^{(0)}(t)=e^{-bt}+b e^{-bt}\int_0^t{\rm d}u\, e^{bu}Q^{(0)}(u)q_d(u)\, ,
\end{equation}
and can be solved directly for all $t$ by defining
\begin{equation}\label{r0q0}
R^{(0)}(t)=e^{bt}Q^{(0)}(t)\,.
\end{equation}
The integral equation for $R^{(0)}(t)$ reads
\begin{equation}
R^{(0)}(t)=1+b\int_0^t{\rm d}u\, R^{(0)}(u)q_d(u)\, .
\end{equation}
Differentiating on both sides gives
\begin{equation}
\frac{{\rm d}R^{(0)}(t)}{R^{(0)}(t)}=bq_d(t){\rm d}t\, ,
\end{equation}
which is integrated as 
\begin{equation}
R^{(0)}(t)=e^{b\int_0^t {\rm d}u\,q_d(u)}\, ,
\end{equation}
after using the initial condition $R_0(0)=1$. Substituting in the definition (\ref{r0q0}), the identity  $Q^{(0)}(t)=e^{-b\int_0^t {\rm d}u[1-q_d(u)]}$ of the main text is obtained.

\section{Mean first passage time}

The expression for the MFPT can be simplified to avoid dealing with three nested integrals,
\begin{eqnarray}
T_{b,d}&=&\int_0^{\infty}{\rm d}t\, q_d(t)e^{-b\int_0^t{\rm d}u[1-q_d(u)]}\\
&=&\int_0^{\infty}{\rm d}t\, e^{-bt} q_d(t)e^{b\int_0^t{\rm d}u\,q_d(u)}
\end{eqnarray}
with $q_d(t)$ given by Eq. (\ref{qdA}). Integrating by parts, one has
\begin{eqnarray}
T_{b,d} &=&\int_0^{\infty}{\rm d}t\, e^{-bt} e^{b\int_0^t{\rm d}u\,q_d(u)}\, -\,\frac{1}{b}\, \\
&=&\int_0^{\infty}{\rm d}t\, e^{-b\int_0^t{\rm d}u\,[1-q_d(u)]}\, -\,\frac{1}{b}\,
\end{eqnarray}
Similarly the integral $\int_0^{t}{\rm d}u\,[1-q_d(u)]$ can be simplified:
\begin{eqnarray}
\int_0^{t}{\rm d}u\,[1-q_d(u)]&=&\int_0^{t}{\rm d}u\,\int_0^{u}{\rm d}y\,P_0(y)e^{-dy}\\
&=&\int_0^{t}{\rm d}u\,e^{-du}(t-u)P_0(u)\,,
\end{eqnarray}
after integrating by parts and grouping the terms. Inserting this last expression into $T_{b,d}$ leads to
\begin{equation}\label{T1genA}
T_{b,d}=\int_0^{\infty} {\rm d}t\,   e^{-b\int_0^t {\rm d}u\, e^{-du}(t-u)P_0(u)}\,-\frac{1}{b}\,.
\end{equation}

\section{Derivation of the joint distribution $P(t,n)$}\label{appC}

When the synthesis process is Poisson, the probability that $n-1$ particles have been created (after the initial one) during the time interval $[0,t]$ and that none of these have crossed the origin at time $t$ is
\begin{equation}
p_{n-1}(t)=\int_0^{t}b{\rm d}t_1\, q_d(t-t_1)
\int_{t_1}^t b{\rm d}t_2\,  q_d(t-t_2)\ldots\int_{t_{n-2}}^t b{\rm d}t_{n-1}\, q_d(t-t_{n-1})\, e^{-bt}\, ,
\end{equation}
where $t_1<t_2<\ldots<t_{n-1}$ are the times of synthesis and the term $e^{-rt}$ ensures that no more particles are created in the union of the intervals $(t_i,t_{i+1})$. One can equivalently relax the order of the $t_i$'s and divide by the number of permutations,
\begin{equation}
p_{n-1}(t)=\left(\int_0^{t}{\rm d}t_1\, q_d(t-t_1)\right)^{n-1}\frac{b^{n-1}e^{-bt}}{(n-1)!}\,.
\end{equation}
We define $P(n,t){\rm d}t$ as the joint probability that $n$ particles have been used up to time $t$ (the first one plus the following $n-1$ ones) and that the first passage at the origin by one of them occurs during the time interval $[t,t+{\rm d}t]$. There are two possibilities: either the first particle is the one that first crosses the origin, or one of the $n-1$ following particles does so. Therefore,
\begin{equation}\label{ptn.1}
P(n,t)=P_0(t)e^{-dt}p_{n-1}(t)+q_d(t)\left(\int_0^tb{\rm d}u\, P_0(t-u)e^{-d(t-u)}\right)p_{n-2}(t)\,.
\end{equation}
In the first term of the rhs of Eq. (\ref{ptn.1}), the factor $e^{-dt}$ is the probability that the first searcher has not degraded before $t$.
The second term tells that the first particle does not find the target [$q_d(t)$], that a certain particle created at a time $u$ finds it, and that the other $n-2$ particles (which are exchangeable) do not find it. After making the variable change $t-u\to u$ and using the relation (\ref{qm}),
\begin{equation}\label{ptn.2}
P(n,t)=P_0(t)e^{-dt}p_{n-1}(t)+q_d(t)[1-q_d(t)]p_{n-2}(t)\,.
\end{equation}
The above expression can be recast in a more compact form as,
\begin{equation}\label{joint2}
P(n,t)=\frac{\partial}{\partial t}\left[
[1-q_d(t)]\left(\int_0^t {\rm d}u\,q_d(u)\right)^{n-1}
\right]\frac{b^{n-1}}{(n-1)!}e^{-bt}\,.
\end{equation}
which is presented in the main text.

One checks that the above expression is normalized. Integrating by part,
\begin{equation}
\int_0^{\infty}{\rm d}t\, P(t,n)=\int_0^{\infty}{\rm d}t\,[1-q_d(t)]\left(\int_0^t {\rm d}u\,q_d(u)\right)^{n-1}\frac{b^{n}}{(n-1)!}e^{-bt}\,.
\end{equation}
Therefore
\begin{eqnarray}
\sum_{n=1}^{\infty}\int_0^{\infty}{\rm d}t\,P(t,n)&=&b\int_0^{\infty}{\rm d}t\, 
[1-q_d(t)]e^{b\int_0^{t}{\rm d}u\,q_d(u)}e^{-bt}\\
&=&b\int_0^{\infty}{\rm d}t\, 
[1-q_d(t)]e^{-b\int_0^{t}{\rm d}u[1-q_d(u)]}\\
&=&1\, ,
\end{eqnarray}
after direct integration.
Similarly, the marginal distribution obtained by summing Eq. (\ref{joint2}) over $n$ coincides with the first passage time distribution, given by $-\partial Q^{(1)}(t)/\partial t$:
\begin{eqnarray}
\sum_{n=1}^{\infty}P(t,n)&=&\frac{\partial}{\partial t}\left[
[1-q_d(t)]e^{b\int_0^{t}{\rm d}u\,q_d(u)}  \right] e^{-bt}\\
&=&\left[-\frac{\partial q_d(t)}{\partial t}+b  [1-q_d(t)]q_d(t) \right]e^{-b\int_0^{t}{\rm d}u[1-\,q_d(u)]} \\
&=&-\frac{\partial}{\partial t}\left[q_d(t) e^{-b\int_0^{t}{\rm d}u[1-\,q_d(u)]}\right]\\
&=&-\frac{\partial Q^{(1)}(t)}{\partial t}\,.
\end{eqnarray}

\section{Calculation of the search cost $\langle n\rangle$}
We recall that 
\begin{equation}\label{q1exact0A}
Q^{(1)}(t)=q_d(t)\, e^{-b\int_0^t {\rm d}u[1-q_d(u)]}\, .
\end{equation}
Denoting $Y_n(t)\equiv[1-q_d(t)]\left(\int_0^t {\rm d}u\,q_d(u)\right)^{n-1}$ for brevity, the mean number of particles used until the reaction time is 
\begin{eqnarray}
\langle n\rangle &=&\sum_{n=1}^{\infty}\int_0^{\infty}{\rm d}t\, nP(t,n)\nonumber\\
&=&\sum_{n=1}^{\infty}\int_0^{\infty}{\rm d}t\, n 
\frac{\partial Y_n(t)}{\partial t}\frac{b^{n-1}}{(n-1)!}e^{-bt}\,\nonumber\\
&=&\frac{\partial }{\partial b_1}\left.
\sum_{n=1}^{\infty}\int_0^{\infty}{\rm d}t\,  
\frac{\partial Y_n(t)}{\partial t}\frac{b_1^{n}}{(n-1)!}e^{-bt}
\right|_{b_1=b}\nonumber\\
&=&\frac{\partial }{\partial b_1}\left.
\sum_{n=1}^{\infty} \int_0^{\infty}{\rm d}t\,  
Y_n(t) \frac{b_1^{n}}{(n-1)!}be^{-bt}
\right|_{b_1=b}\,.
\end{eqnarray}
We further have
\begin{eqnarray}
\langle n\rangle&=&b\frac{\partial }{\partial b_1}\left.
\int_0^{\infty}{\rm d}t\,  [1-q_d(t)]\sum_{n=1}^{\infty}\left(\frac{\left(b_1\int_0^t {\rm d}u\,q_d(u)\right)^{n-1}} {(n-1)!}\right)b_1e^{-bt}
\right|_{b_1=b}\nonumber\\
&=&b\int_0^{\infty}{\rm d}t\, e^{-bt}[1-q_d(t)]
\frac{\partial }{\partial b_1}\left.\left[ 
b_1e^{b_1\int_0^t {\rm d}u\,q_d(u)}
\right]\right|_{b_1=b}\nonumber\\
&=&b\int_0^{\infty}{\rm d}t\, [1-q_d(t)]\left[1+b\int_0^t {\rm d}u\,q_d(u)\right]e^{-b\int_0^t {\rm d}u\,[1-q_d(u)}\nonumber\\
&=&1+b^2\int_0^{\infty}{\rm d}t\, [1-q_d(t)]\left(\int_0^t {\rm d}u\,q_d(u)\right)e^{-b\int_0^t {\rm d}u\,[1-q_d(u)]}\nonumber\\
&\ &{\rm (after\ direct\ integration\ of\ the\ first\  term\ in\  [\ldots]\ above)}\nonumber\\
&=&1+b \int_0^{\infty}{\rm d}t\, q_d(t)e^{-b\int_0^t {\rm d}u\,[1-q_d(u)]}\quad {\rm (after\ inegrating\ by\ parts)}\nonumber\\
&=& 1+b \int_0^{\infty}{\rm d}t\, Q^{(1)}(t)\quad ({\rm from\ Eq.\ (\ref{q1exact0A})})\nonumber\\
&=&1+bT_{b,d}\, .
\end{eqnarray}

\section{Expansion  of the MFPT $T_{b,d}$ at small $b$ and $d$}\label{appB}

Instead of expanding the exact solution, 
\begin{equation}
T_{b,d}=\int_0^{\infty} {\rm d}t\,   e^{-b\int_0^t {\rm d}u\, e^{-du}P_0(u)(t-u)}\,-\frac{1}{b}\,,
\end{equation}
at small $b$ and $d$ (both assumed to be of same order), we can start from the identity $Q^{(1)}(t)=q_d(t)Q^{(0)}(t)$ and Eq. (\ref{renewq0A}) to obtain the \lq\lq renewal-like" equation for $Q^{(1)}(t)$,
\begin{equation}\label{q1renew}
Q^{(1)}(t)=q_d(t)e^{-bt}+bq_d(t)\int_0^{t}{\rm d} t'\, e^{-b(t-t')}Q^{(1)}(t')\, .
\end{equation}
When $b$ is small, the second term of the rhs of Eq. (\ref{q1renew}) can be treated as a perturbation of the first one. Therefore one can substitute $Q^{(1)}(t')$ in the integral by its leading behavior 
$q_d(t')e^{-bt'}$,
\begin{equation}\label{q1renew.2}
Q^{(1)}(t)=q_d(t)e^{-bt}+bq_d(t)e^{-bt}\int_0^{t}{\rm d} t'\, q_d(t')+{\rm h.o.t.}\, ,
\end{equation}
where h.o.t. represents higher order terms whose magnitude will be evaluated later. The MFPT is obtained from Eq. (\ref{q1renew.2}) by using the general relation $T_{b,d}=\int_0^{\infty}{\rm d}t\,Q^{(1)}(t)$,
\begin{equation}\label{t1pert.1}
T_{b,d}=\tilde{q}_d(b)+\frac{b^2}{2}\int_0^{\infty}{\rm d}t\, e^{-bt}\left(\int_0^t{\rm d}t'\,q_d(t')\right)^2+{\rm h.o.t.}\,, 
\end{equation}
after integration by parts of the second term of the rhs. [The same expression can be obtained starting from the exact solution Eq. (\ref{q1exact0A}) and writing $e^{b\int_0^{t}{\rm d}u\,q_d(u)}=\sum_{n=0}^{\infty}\frac{(b\int_0^{t}{\rm d}u\,q_d(u))^n}{n!}$ and retaining the first 3 terms of the sum.] 

Taking the Laplace transform of $q_d(t)$, which is given by 
\begin{eqnarray}\label{qm}
    q_{d}(t)=1-\int_0^t d\tau e^{-d\tau}P_0(\tau)\, ,
\end{eqnarray} 
the first term of
Eq. (\ref{t1pert.1}) reads
\begin{equation}
\tilde{q}_d(b)=\frac{1-\tilde{P}_0(b+d)}{b}\, .
\end{equation}
At small $b$ and $d$,
\begin{equation}\label{firstterm}
\tilde{P}_0(b+d)=1-(b+d)\langle T_0\rangle+\frac{(b+d)^2}{2}\langle T_0^2\rangle+\ldots\, ,
\end{equation}
hence
\begin{equation}\label{qmpert}
\tilde{q}_d(b)=\frac{b+d}{b}\langle T_0\rangle-\frac{(b+d)^2}{2b}\langle T_0^2\rangle+\ldots\,,
\end{equation}
where the neglected terms are quadratic, i.e., of order $b^2$, $d^2$ and $bd$.

To evaluate the second term of the rhs of Eq. (\ref{t1pert.1}), let us first rewrite $q_d(t)$ as
\begin{equation}\label{qm2}
q_d(t)=\int_t^{\infty}{\rm d}t'P_0(t')+\int_0^t{\rm d}t'(1-e^{-dt'})P_0(t')\,,
\end{equation}
from the normalization of $P_0(t')$ to unity. Recall that $P_0(t)$ is assumed to rapidly decay to $0$ with $t$ (say, exponentially) over a characteristic time-scale $1/\mu$. Since $d$ is small, i.e., $1/\mu\ll 1/d$, the factor $1-e^{-dt'}$ in Eq. (\ref{qm2}) can be replaced by  $dt'$ at first order, for all values of $t'$. Hence,
\begin{eqnarray}\label{qm3}
q_d(t)&=&\int_t^{\infty}{\rm d}t'P_0(t')+d\int_0^t{\rm d}t'\,t'P_0(t')+\ldots \nonumber\\
&=& q_0(t)+d\int_0^t{\rm d}t'\,t'P_0(t')+\ldots\, ,
\end{eqnarray}
where $q_0(t)$ is identified with the survival probability of the underlying process with $d=0$. With Eq. (\ref{qm3}), one deduces 
\begin{equation}\label{intqmt}
\int_0^t{\rm d}t'q_d(t')=\int_0^{t}{\rm d}t' q_0(t') + td\int_0^t{\rm d}t't' P_0(t')-d\int_0^t{\rm d}t' t'^2 P_0(t')+\ldots\,
\end{equation}
after an integration by parts of the last term of Eq. (\ref{qm3}).
The second term of the rhs of Eq. 
(\ref{t1pert.1})
can be decomposed into two parts. We define
\begin{equation}
I=\int_0^{\infty}{\rm d}t\, e^{-bt}\left(\int_0^t{\rm d}t'\,q_d(t')\right)^2=I_1+I_2\, ,
\end{equation}
by writing $\int_0^{\infty}{\rm d}t=\int_0^{A/\mu}{\rm d}t+\int_{A/\mu}^{\infty}{\rm d}t$, with $A$ a fixed constant much larger than 1. 

In the integral $I_2$, defined as 
\begin{equation}\label{i2}
I_2=\int_{A/\mu}^{\infty}{\rm d}t\, e^{-bt}\left(\int_0^t{\rm d}t'\,q_d(t')\right)^2\,,
\end{equation}
$t$ is much larger than $1/\mu$. Consequently, in the expression (\ref{intqmt}) for the quantity $\int_0^t{\rm d}t'\,q_d(t')$, one can replace the upper integration bound $t$ of the three integrals by $\infty$, which gives
\begin{equation}\label{intqmi2}
\int_0^t{\rm d}t'q_d(t')=\langle T_0\rangle+td\langle T_0\rangle-d\langle T_0^2\rangle
+\ldots\,,
\end{equation}
where the neglected terms are ${\cal O}(d^2)$. Substituting Eq. (\ref{intqmi2}) into (\ref{i2}) and grouping the factors $e^{-bt}t^0$, $e^{-bt}t$ and $e^{-bt}t^2$ in the integrand, one obtains, respectively, 
\begin{equation}
I_2=I_2^{(a)}+I_2^{(b)}+I_2^{(c)}\,.
\end{equation}
with 
\begin{eqnarray}
I_2^{(a)}&=&\left(\langle T_0\rangle-d\langle T_0^2\rangle\right)^2\int_{A/\mu}^{\infty}{\rm d}t\,e^{-bt}\simeq \frac{\langle T_0\rangle^2}{b}(1-e^{-bA/\mu})\simeq \frac{\langle T_0\rangle^2}{b}+{\cal O}(b^0)\label{i2a}\\
I_2^{(b)}&=&2d\langle T_0\rangle\left(\langle T_0\rangle-d\langle T_0^2\rangle \right)\int_{A/\mu}^{\infty}{\rm d}t\,e^{-bt} t
\simeq \frac{2d^2\langle T_0\rangle^2}{b^3}+{\cal O}(b^0)\label{i2b}\\
I_2^{(c)}&=&d^2\langle T_0\rangle^2 \int_{A/\mu}^{\infty}{\rm d}t\,e^{-bt} t^2\simeq \frac{2d^2\langle T_0\rangle^2}{b^3}+{\cal O}(b^0)\label{i2c}\, ,
\end{eqnarray}
assuming that $b$ and $d$ are of the same order. 

The other part $I_1$ of the integral $I$ is given by
\begin{equation}\label{i1}
I_1=\int_0^{A/\mu}{\rm d}t\, e^{-bt}\left(\int_0^t{\rm d}t'\,q_d(t')\right)^2\,.
\end{equation}
Since $q_d(t)$ is bounded by $1$,
\begin{equation}
I_1<\int_0^{A/\mu}{\rm d}t\, e^{-bt}t^2=\frac{1}{3}\left(\frac{A}{\mu}\right)^3+{\cal O}(b)\,,
\end{equation}
therefore $I_1\ll I_2$ as the latter is of order $1/b$ at leading order, see Eqs. (\ref{i2a})-(\ref{i2c}). Gathering Eqs. (\ref{i2a})-(\ref{i2c}) one obtains the second contribution to $T_{b,d}$ in Eq. (\ref{t1pert.1}),
\begin{eqnarray}
\frac{b^2}{2}\int_0^{\infty}{\rm d}t\, e^{-bt}\left(\int_0^t{\rm d}t'\,q_d(t')\right)^2&=&\langle T_0\rangle^2\left(\frac{b}{2}+\frac{d^2}{b}+d\right)+{\cal O}(b^2)\nonumber\\
&=&\langle T_0\rangle^2\frac{(b+d)^2+d^2}{2b}+{\cal O}(b^2)\, .\label{ipert}
\end{eqnarray}
Adding the two contributions (\ref{qmpert}) and (\ref{ipert}) one obtains
\begin{equation}
T_{b,d}=\frac{b+d}{b}\langle T_0\rangle-\frac{(b+d)^2}{2b}\langle T_0^2\rangle+\langle T_0\rangle^2\,\frac{(b+d)^2+d^2}{2b}+{\cal O}(b^2)\,,
\end{equation}
which is can be rewritten as
\begin{equation}\label{T1pertA}
T_{b,d}=\frac{b+d}{b}\langle T_0\rangle-\frac{(b+d)^2}{2b} \langle T_0\rangle^2 \left[CV^2-\frac{d^2}{(b+d)^2}\right]+ {\rm h.o.t.}\,,
\end{equation}
as presented in the main text.

\section{MFPT for Brownian particles under SSD in an interval}
The probability density $\rho(x,t)$ for a Brownian particle diffusing in a finite interval with absorbing boundaries at $x=0$ and $x=L$ is given by (see, e.g., Ref. [50] of the main text)
\begin{equation}
    \rho_0(x,t)=\frac{2}{L}\sum^{\infty}_{n=1}\exp\left(-\frac{\pi^2 n^2Dt}{L^2}\right)\sin\left(\frac{n\pi x_0}{L}\right)\sin\left(\frac{n\pi x}{L}\right)\, .
\end{equation}
Therefore, the survival probability for this process will be given by
\begin{equation}
    q_0(t)=\int_{0}^L {\rm d}x\ \rho_0(x,t)= \frac{4}{\pi}\sum^{\infty}_{n=1}\exp\left(-\frac{\pi^2 (2n-1)^2Dt}{L^2}\right)\sin\left(\frac{(2n-1)\pi x_0}{L}\right)\frac{1}{2n-1}
\end{equation}
The PDF of the exit time, or first passage at one of the two absorbing walls, is deduced
\begin{equation}
    P_0(t)=-\frac{\partial q_0(t)}{\partial t}= \frac{4\pi D}{L^2}\sum^{\infty}_{n=1}(2n-1)\exp\left(-\frac{\pi^2 (2n-1)^2Dt}{L^2}\right)\sin\left(\frac{(2n-1)\pi x_0}{L}\right).
\end{equation}
We then plug this expression into the general formula (\ref{T1genA}) for the MFPT of the SSD process.
The integral $I(t)=\int_0^t {\rm d}u\ e^{-du} (t-u) P_0(u)$ in Eq. \eqref{T1genA} can be computed as
\begin{eqnarray}
    I(t)&=&\frac{4\pi D}{L^2}\sum^{\infty}_{n=1}(2n-1)\sin\left(\frac{(2n-1)\pi x_0}{L}\right)\int_0^t {\rm d}u\ (t-u) e^{-\sigma_n u}\nonumber\\
    &=&\frac{4\pi D}{L^2}\sum^{\infty}_{n=1}(2n-1)\sin\left(\frac{(2n-1)\pi x_0}{L}\right)\frac{e^{-\sigma_n t}+\sigma_n t-1}{\sigma_n^2}
\end{eqnarray}
where we have defined $\sigma_n=d+\frac{\pi^2 (2n-1)^2D}{L^2}$.
We can now write the mean first passage time as
\begin{eqnarray}
    T_{b,d}=\int_0^\infty{\rm d}t\ \prod_{n=1}^{\infty}\exp\left[-\frac{4b\pi D(2n-1)}{L^2}\sin\left(\frac{(2n-1)\pi x_0}{L}\right)\frac{e^{-\sigma_n t}+\sigma_n t-1}{\sigma_n^2}\right]-\frac{1}{b}\, ,
\end{eqnarray}
which is evaluated numerically by the trapezium method.  Taking the product with $n$ up to $100$ is sufficient for a good convergence. 

\end{document}